# A Cyber-Physical Perspective to Pinning-Decision for Distributed Multi-Agent Control in Microgrid against Stochastic Communication Disruptions

[1]Samson S. Yu, *Member, IEEE,* and [2]Tat Kei Chau, *Member, IEEE*

*Abstract*—In this study, we propose a decision-making strategy for pinning-based distributed multi-agent (PDMA) automatic generation control (AGC) in islanded microgrids against stochastic communication disruptions. The target microgrid is construed as a cyber-physical system, wherein the physical microgrid is modeled as an inverter-interfaced autonomous grid with detailed system dynamic formulation, and the communication network topology is regarded as a cyber-system independent of its physical connection. The primal goal of the proposed method is to decide the minimum number of generators to be pinned and their identities amongst all distributed generators (DGs). The pinning-decisions are made based on complex network theories using the genetic algorithm (GA), for the purpose of synchronizing and regulating the frequencies and voltages of all generator busbars in a PDMA control structure, i.e., without resorting to a central AGC agent. Thereafter, the mapping of cyber-system topology and the pinning decision is constructed using deep-learning (DL) technique, so that the pinning-decision can be made nearly instantly upon detecting a new cyber-system topology after stochastic communication disruptions. The proposed decision-making approach is verified using a 10-generator, 38-bus microgrid through time-domain simulation for transient stability analysis.

## I. INTRODUCTION

With the electric power systems transforming into a more sustainable state, an increasing penetration of renewable energy is being integrated into the electric grid. This leads to a burgeoning number of distributed generators that are boosting the transformation of the traditional centralized electricity generation and transmission architecture into a more distributed one. In any power system, automatic generation control is an essential control mechanism which regulates the frequency and power flow between two adjacent areas for reliable power grid operations [1]. Traditionally, AGC is achieved through central control agents governed by power system operators within their jurisdiction, through directly controlling large-scale fuel-based power generators to produce more or less electricity against various power system conditions. However, as the scale of microgrids comprising of DGs grows rapidly, a central AGC agent that regulates all small-scale distributed generators to achieve intelligent power system control is no longer viable due to the colossal communication requirement [2]. This has motivated the advent of the multi-agent system (MAS) with its applications in power system control, which mainly include economic dispatch [3], heterogeneous storage systems [4] and power system control [5], [6]. The overarching improvement for MAS, in comparison to the conventional centralized AGC structure, is that the control agents, or generators, only have access to the information from a limited number of neighbors, substantially reducing the requirement for communication demanded by a central AGC [7]. In an MAS setup, there could still be a central agent able to pass and receive timely power system information, including system time and system events, to and from control agents without excessive data transmission and processing, so as to improve situational awareness of the system [8].

For automatic generation control in microgrids, pinning-based distributed MAS was recently proposed and has gained popularity among researchers due to its proven feasibility and easy implementation. The PDMA control is generally conducted at a secondary control level through autonomously changing the nominal frequencies and voltages of control agents, based on the droop control theory [5]. Upon system disturbances, such as load demand variations, selected control agents with given communication topology will alter their nominal frequencies and voltages via an automated control mechanism to realize frequency and voltage synchronization (or consensus) and restoration within the microgrid. Another main requirement in AGC is that the total load demand and power loss need to be shared by all generators based on their capacities and droop coefficients [9], which however were not considered in [5]. In addition, the rationale of pinning decisions, i.e., the number of generators that need to be pinned and what they are, has been absent in most pinning-control research [5], [10], [11], where the pinning of generators were performed on a trial-and-error basis, until a year ago when the authors in [6] employed complex network theories to corroborate the choice of the pinning set. Notwithstanding its merit and vigorous mathematical derivations of the control methodology, the most recent pertinent work in [6] bears certain shortcomings: the studied microgrid was oversimplified without considering power flow, power loss and load model in the microgrid of interest, which can not fully represent the system characteristics from an electrical perspective. The communication was deemed identical to the physical topology via power-line communication technique, which limits its applicability in contemporary electric grids; with an increasing number of wireless data transmission techniques, such phasor measurement units (PMUs), the physical and communication systems have become totally separable. Lastly, the convergence rate of the decision-making method was computed purely based on cyber-system topology, failing to address the electrical features of microgrids.

[1]S. Yu is with the School of Engineering, Deakin University, Melbourne, Australia (email: s.yu@ieee.org). [2]T. Chau is with the School of Engineering, University of Western Australia, Perth, Australia .

The above important problems have not yet been addressed in the literature, and to the best of our knowledge, they need to be appropriately understood and resolved from a cyber-physical perspective. Due to the dramatic development and deployment of wireless measuring and data transmission devices and techniques, interpreting a modern power system as a cyber-physical system is a necessary consideration, which however requires substantially more understanding and investigations into both areas and seamlessly integrate them. In a very recent study [12], the authors incorporated complex network theories into complex power systems with fixed physical connection but various communication topologies to investigate the vulnerability of the power system against cascading failure. Similar ways of perceiving a power system have only been considered in recent couple of years [13]–[15], where small signal stability analysis, line failure detection, and critical modes were studied respectively for large-scale power systems. These pieces of research work have not been applied to inverter-interfaced microgrids, which have a totally different modeling approach, nor a multi-agent distributed control structure.

Therefore, pinning-based secondary control needs to be further developed, with simultaneous consideration of the physical and cyber systems that collectively represent modern microgrids, to improve the robustness of the control system upon stochastic cyberattacks that may disrupt communication channels. In this paper, from a cyber-physical perspective, we propose a decision-making strategy for PDMA that has high practicality and implementability for microgrid research. To achieve this, we first utilize an improved static model of islanded microgrids documented in [16] and dynamic model for inverter-interfaced microgrids in [17], to formulate a complete 10-generator 38-bus autonomous microgrid. Assuming communication connections are completely independent of the physical architecture, a random communication topology is generated as a typical small-world connection. We then use this initial communication topology as the base system and simulate stochastic communication disruptions until communication sub-space starts to appear. Based on complex network theories and frequency and voltage characteristics, an optimization problem, for maximizing the CDR of the microgrid, is formed for a given communication topology, to solve for the minimum number and identities of DGs to be pinned. Thereafter, time-domain simulations for the 10-generator, 38-bus microgrid are performed to demonstrate the functionality of the proposed method. Lastly, to substantially reduce the computational burden of the proposed PDMA when operating on a real-time basis, a DL-based pinning-decision mechanism is devised, enabling fast real-time pinning-decision for the microgrid of interest.

## II. Modeling of a Multi-Inverter Microgrid

In this section, descriptions of the mathematical model of each electrical component in a microgrid are presented, including the inverter, network, load and power flow, under the PDMA control frame.

### A. DG Inverter with Droop Control

A typical DG-inverter structure comprises a power source, voltage source converters, an LCL filter and internal controllers for the power converters. To simplify the study without losing generality, we assume the power source is able to produce the required amount of power for the islanded microgrid under both steady-state operation and transient operation caused by small disturbances. In this paper, a multiple-stage control loop is illustrated in Fig. 1. This control scheme consists of three controllers–the power, voltage and current controller.

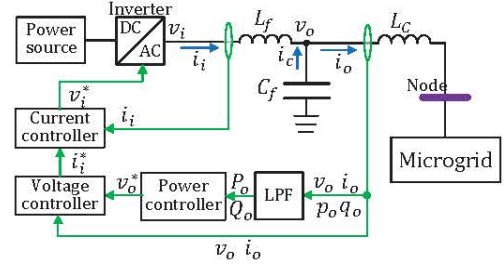

Fig. 1: DG inverter schematic

*1) Power Controller:* Voltage and frequency references are generated by the power controller based on the filtered local active, reactive power, current and voltage measurements from the LCL filter. As shown in Fig. 1, the calculated instantaneous active and reactive output power, $\tilde{p}_o$ and $\tilde{q}_o$, passes through a low pass filter (LPF) to obtain the power quantities corresponding to the fundamental components, $P_o$ and $Q_o$, using the following equations,

$$\tilde{p}_o = i_o^d v_o^d + i_o^q v_o^q, \qquad \tilde{q}_o = i_o^d v_o^q - i_o^q v_o^d, \qquad (1)$$
$$\dot{P}_o = \omega_c (\tilde{p}_o - P_o), \qquad \dot{Q}_o = \omega_c (\tilde{q}_o - Q_o), \qquad (2)$$

where $v_o^d$, $v_o^q$, $i_o^d$ and $i_o^q$ are the instantaneous $d-q$ voltage and current, and $\omega_c$ is the cut-off frequency of the LPF filter.

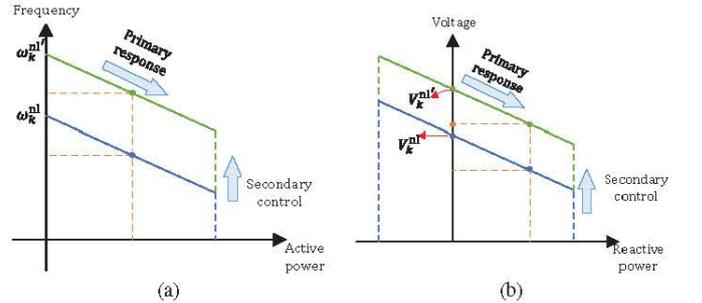

Fig. 2: $\omega - P$ droop and $V - Q$ droop

Droop control is adopted to regulate the power sharing between DGs. As shown in Fig. 2, a change in power generation by the inverter-based power sources, which will alter the inverter frequency and voltage, based on the following relations,

$$\omega_k = \omega_k^{nl} - m_{p_k} P_{G_k}, \qquad V_k = V_k^{nl} - n_{q_k} Q_{G_k}, \qquad (3)$$

$$v_{o_k}^{*d} = \left[(V_k \cos\theta_k + r_c i_{o_k}^d - \omega_k L_c i_{o_k}^q)^2 \right.$$
$$\left. + (V_k \sin\theta_k - r_c i_{o_k}^q + \omega_k L_c i_{o_k}^d)^2 \right]^{1/2},$$
$$v_{o_k}^{*q} = 0, \qquad (4)$$

where subscript $k$ is used to denote the number of node and superscript $*$ represents the reference value to be used in subsequent control loops. Term $\omega_k$ is the inverter frequency, $\omega_k^n$ and $V_k^n$ are the nominal frequency and voltage, $m_{p_k}$ and $n_{q_k}$ are the $\omega - P$ and $V - Q$ droop coefficients, and $\theta_k$ is the phase angle of the inverter voltage. The pinning-based AGC control is performed through changing the nominal frequency and voltage of DGs, e.g., $\omega_k^n$ and $V_k^n$ move to $\omega_k^{n'}$ and $V_k^{n'}$ respectively in Fig. 2.

*2) Current and Voltage Control:* As shown in Fig. 3, both current and voltage controllers consist of two standard PI-controllers to regulate the $d - q$ components of voltage and current. As mentioned earlier, reference values of the $d$ and $q$-axis output voltage $v_{o_k}^{*d}$ and $v_{o_k}^{*q}$ are generated by the power controller, and the outputs of the PI-controllers in the voltage controller are then added together with feed-forward terms $K_F i_{o_k}^d$ and $K_F i_{o_k}^q$, to generate the references of the inverter currents $i_{i_k}^{*d}$ and $i_{i_k}^{*q}$, where $K_F$ is the feed-forward gain and $i_{o_k}^d$ and $i_{o_k}^q$ are the $d - q$ components of the output current of the LCL filter as shown in Fig. 1. Similarly, the generated current references are fed into current controller to generate the reference values for the inverter voltage $v_{i_k}^{*d}$ and $v_{i_k}^{*q}$ with the similar approach used in the voltage controller. Detailed mathematical expressions can be found in [17].

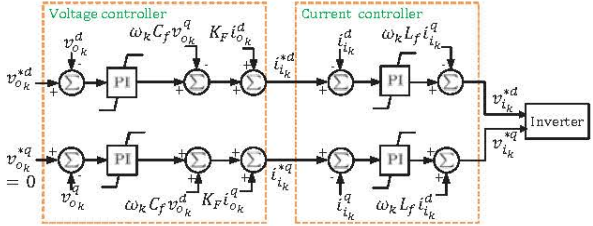

Fig. 3: Current and voltage controllers in DG inverter

*B. LCL Filter*

The following differential equations summarizes the dynamics of the output LCL filter connected to an inverter (Fig. 5 shows the topology),

$$dv_{o_k}^{d,q}/dt = 1/C_f \left(i_{i_k}^{d,q} - i_{o_k}^{d,q} + \omega_k C_f v_{o_k}^{q,d}\right),$$
$$di_{i_k}^{d,q}/dt = 1/L_f \left(v_{i_k}^{d,q} - v_{o_k}^{d,q} - r_f i_{i_k}^{d,q} + \omega_k L_f i_{i_k}^{q,d}\right),$$
$$di_{o_k}^{d,q}/dt = 1/L_c \left(v_{o_k}^{d,q} - v_k^{d,q} - r_c i_{o_k}^{d,q} + \omega_k L_c i_{o_k}^{q,d}\right), \quad (5)$$

where $v_k^{d,q}$ is the $d - q$ components of voltage at $k$th node, $L_c$ is the coupling inductance in the LCL filter, and $r_f$ and $r_c$ are the parasitic resistances of the filtering coupling inductors respectively.

*C. Load modeling and Network Equations*

Instead of assuming constant power consumption or impedances as in large-scale power system studies [18], [19], in this study, electric loads and transmission lines are modeled as $R - L$ impedances, which vary with system frequency. The following equation describes the relation between current and voltage of transmission lines,

$$di_{\text{li}_{k_j}}^{D,Q}/dt =$$
$$1/L_{\text{li}_{k_j}} \left(v_k^{D,Q} - v_j^{D,Q} - R_{\text{li}_{k_j}} i_{\text{li}_{k_j}}^{D,Q} + \omega L_{\text{line}_{k_j}} i_{\text{line}_{k_j}}^{Q,D}\right), \quad (6)$$

where $L_{\text{li}_{k_j}}$ and $R_{\text{li}_{k_j}}$ are the inductance and resistance of the transmission line connecting node $k$ and $k$, $i_{\text{line}_{k_j}}$ is the current flowing from node $k$ to node $k$, $v_k$ is the voltage of node $k$, and $D - Q$ represents the direct and quadrature components of the common reference frame (which differs from the inverter $d - q$ reference frame for each DG bus) in the islanded microgrid.

The following equations describe the relations between current and voltage at load nodes,

$$di_{\text{load}_k}^{D,Q}/dt =$$
$$1/L_{\text{load}_k} \left(v_k^{D,Q} - R_{\text{load}_k} i_{\text{load}_k}^{D,Q} + \omega L_{\text{load}_k} i_{\text{load}_k}^{Q,D}\right), \quad (7)$$

where $L_{\text{load}_k}$ and $R_{\text{load}_k}$ are the inductance and resistance of the load connected to node $k$, and $i_{\text{load}_k}$ is the current flowing into node $k$. In microgrid modeling, the current and voltage at each node also have the following relation,

$$v_k^{D,Q} R_N^{-1} = i_{o_j}^{D,Q} - i_{\text{load}_j}^{D,Q} + \sum_{k=1, k \neq j}^{N} i_{\text{line}_{k_j}}^{D,Q}, \quad (8)$$

where $R_N$ is the virtual resistance connecting each node to the ground. The introduction of the virtual resistance is to ensure the numerical stability when conducting the simulation experiments for the microgrid [17].

## III. Small Signal Stability Analysis Model

In this section, a brief discussion on SSSA model formulation will be presented, where a Differential Algebraic Equation (DAE) formulation is used, and system linearization is performed to realize the SSSA for microgrids.

The nonlinear mathematical model describing a microgrid can be written in the following compact form [1], [17],

$$d\mathbf{X}/dt = f(\mathbf{X}, \mathbf{\Upsilon}, \mathbf{V}, \mathbf{U}),$$
$$0 = g_1(\mathbf{X}, \mathbf{\Upsilon}, \mathbf{V}),$$
$$0 = g_2(\mathbf{X}, \mathbf{V}), \quad (9)$$

where $\mathbf{X}$ is the dynamic state vector of the microgrid, including the dynamics of the inverters, DGs, controllers, transmission lines and loads; $\mathbf{V}$ is the vector containing the voltage magnitudes and angles of all nodes; $\mathbf{\Upsilon}$ represents inverter algebraic variables; and $\mathbf{U}$ is the control input vector comprising droop coefficients and nominal frequency and voltage of the microgrid. Function $f(\cdot)$ is the system state-space function, $g_1(\cdot) = 0$ is the inverter algebraic equation set, and $g_2(\cdot) = 0$ is the network equation set.

## A. Improved Power Flow Analysis

The steady-state solution of the islanded droop-based microgrid can be obtained by power flow analysis as studied [16] using Newton-Raphson method. However, when pinning-based secondary frequency and voltage strategies are deployed in an islanded microgrid, its steady-state frequency settles to a preset set-point programmed in the secondary controller instead of free running according solely to the droop characteristics described in (3). Therefore, it is necessary to modify and improve the algorithm in order to obtain the power flow solution when implementing a PDMA control structure. A brief mathematical description is summarized as follows,

$$W_{i+1} = J_i^{-1} \Delta \mathcal{P}_i + \Delta W_i, \quad (10)$$

where $W$, $\Delta \mathcal{P}$ and $\Delta W$ are the modified unknown vector, power mismatch vector and correction vector at $i^{\text{th}}$ iteration respectively, and $J$ is the Jacobian matrix of the power flow formulation. To demonstrate the islanding feature, the modified unknowns of the power flow problem are given as follows,

$$W = \begin{bmatrix} \omega^{\text{nl}} & V^{\text{nl}} & \theta & |V_{\text{load}}| \end{bmatrix}^T, \quad (11)$$

where $\theta$ and $V$ are respectively the voltage phase angles and magnitudes for all nodes as stated in original formulation of a power flow problem [1], and $\omega^{\text{nl}}$ and $V^{\text{nl}}$ are two vectors that contain the nominal frequency and voltage for each inverter node with the secondary control mentioned in Section II-A. The microgrid will have known unanimous frequency $\omega$ and voltage level $V$ at every DG bus-bar. The formulation of the modified power mismatch vector $\Delta P$ can be found in [16]. The dimensions of $W$, $\omega^{\text{nl}}$, $V^{\text{nl}}$, $\theta$, $|V_{\text{load}}|$ for an $m$-DG $N$-bus microgrid are respectively $(2N+m-1) \times 1$, $m \times 1$, $m \times 1$, $(N-1) \times 1$, and $(N-m) \times 1$, under the premises that the angular reference bus is a generator bus with a known voltage level $V$ and $\theta_{\text{ref}} = 0$, which is a load bus. Note that the power generation of each inverter node has now become,

$$P_{G_k} = \frac{1}{m_{p_k}} \left( \omega_k^{\text{nl}} - \omega_k \right), Q_{G_k} = \frac{1}{n_{q_k}} \left( V_k^{\text{nl}} - V_k \right), \quad (12)$$

and $R_N$ is modeled as a shunt impedance. Corresponding modifications need to be made when computing the Jacobian matrix, which are shown as follows,

$$\Delta \mathcal{P} = \begin{bmatrix} \Delta P_a & \Delta Q_a & \Delta P & \Delta Q & \Delta \Theta \end{bmatrix}^T, \quad (13)$$

where the dimension for each element is $(2N+m-1) \times 1$, $1 \times 1$, $1 \times 1$, $(N-1) \times 1$, $(N-1) \times 1$ and $(m-1) \times 1$, respectively, and

$$P_a = P_{\text{tot}} - P_{\text{sys}}, \quad Q_a = Q_{\text{tot}} - Q_{\text{sys}}, \quad (14)$$

$$P_{\text{tot}} = P_{\text{load}} + P_{\text{loss}}, \quad Q_{\text{tot}} = Q_{\text{load}} + Q_{\text{loss}}, \quad (15)$$

$$\Delta P = P - P_{\text{cal}}, \quad \Delta Q = Q - Q_{\text{cal}}, \quad (16)$$

$$\Delta \Theta = m_{p_1} P_{G_1} I^{(m-1) \times 1} - [m_{p_2} P_{G_2}, \cdots, m_{p_m} P_{G_m}]^T (17)$$

$$P_{G_k} = \frac{1}{m_{p_k}} \left( \omega_k^{\text{nl}} - \omega_k \right), \quad Q_{G_k} = \frac{1}{n_{q_k}} \left( V_k^{\text{nl}} - V_k \right), \quad (18)$$

$$P_{\text{cal}_k} = V_k \sum_{j=1}^{N} V_j |Y_{(k,j)}| \cos \left( \theta_k - \theta_j - \gamma_{jk} \right),$$

$$Q_{\text{cal}_k} = V_k \sum_{j=1}^{N} V_j |Y_{(k,j)}| \sin \left( \theta_k - \theta_j - \gamma_{jk} \right), \quad (19)$$

$$P_{\text{sys}} = \sum_{k=1}^{m} P_{G_k} = \sum_{k=1}^{m} \frac{1}{m_{p_k}} \left( \omega_k^{\text{nl}} - \omega_k \right),$$

$$Q_{\text{sys}} = \sum_{k=1}^{m} Q_{G_k} = \sum_{k=1}^{m} \frac{1}{n_{q_k}} \left( V_k^{\text{nl}} - V_k \right), \quad (20)$$

$$P_{\text{loss}} = \frac{1}{2} \sum_{k=1}^{N} \sum_{j=1}^{N} \Re \left\{ Y_{(k,j)} \left( V_k^* V_j + V_k V_j^* \right) \right\},$$

$$Q_{\text{loss}} = -\frac{1}{2} \sum_{k=1}^{N} \sum_{j=1}^{N} \Im \left\{ Y_{(k,j)} \left( V_k^* V_j + V_k V_j^* \right) \right\}, \quad (21)$$

where the extra equation set (17) is introduced to achieve appropriate power sharing based on droop characteristics, which will be mentioned again in Section IV. After solving the power flow problem, steady-state $X^{\text{ss}}$ can be obtained, from which the initial condition of voltage magnitudes and angles of all nodes $V^{\text{ss}}$ can be extracted.

## IV. PINNING-BASED DISTRIBUTED MULTI-AGENT AUTOMATIC GENERATION CONTROL

### A. PDMA Physical Control

As mentioned in earlier sections, the two main aspects in PDMA formulation is (1) the physical droop-based secondary control and (2) the communication network and information exchange. In this study, we assume that all DGs have droop-based frequency and voltage control mechanisms as stated in II, with the control variables being the nominal frequency $\omega_k^n$ and voltage $V_k^n$ ($k = 1, 2, \cdots, m$) for each DG. The information exchange between two cyber-connected DGs includes frequency, voltage and active power. The following equation shows the control of nominal voltage at DG bus $k$,

$$V_k^{\text{nl}} = n_{q_k} Q_{G_k} + C_v \int \left( \sum_{j=1}^{m} A_{\text{adj}(j,k)} (V_j - V_k) \right.$$
$$\left. + \Psi_{(k,k)} C_{gv(k,1)} (V_{\text{ref}} - V_k) \right), \quad (22)$$

where $C_v$ is the voltage control gain, $A_{\text{adj}}$ is the adjacency matrix of the communication network (which can be calculated based on [20]), $C_{gv}$ is the voltage pinning gain vector, and $\Psi$ is the pinning matrix, which is defined as follows,

$$\Psi = \begin{bmatrix} \psi_1 & 0 & \cdots & 0 \\ 0 & \psi_2 & \cdots & 0 \\ \vdots & \vdots & \ddots & \vdots \\ 0 & 0 & \cdots & \psi_m \end{bmatrix}, \quad (23)$$

where $\psi$ is a Boolean variable ($= 1$ or $0$) for pinning or not pinning a particular DG.

Similarly, the nominal frequency of DG at bus $k$ can be controlled as,

$$\omega_k^{\text{nl}} = m_{p_k} P_{G_k} + C_\omega \int \left( \sum_{j=1}^{m} A_{\text{adj}(j,k)} (\omega_j - \omega_k) \right.$$
$$\left. + \Psi_{(k,k)} C_{g\omega(k,1)} (\omega_{\text{ref}} - \omega_k) \right) + C_P \int \Bigg($$
$$\sum_{j=1}^{m} A_{\text{adj}(j,k)} \left( m_{p_k} P_{G_k} - m_{p_j} P_{G_j} \right) \Bigg), \quad (24)$$

where the last addition term is to share the power demand among all DGs based their droop characteristics, $C_\omega$ is the frequency pinning gain, $C_P$ is the power sharing balancing gain and $C_{g\omega}$ is the frequency pinning gain.

*B. PDMA Pinning Decision Formulation based on Complex Network Theory*

As stated before, the pinned generator has a three-fold task– (a) synchronizing its frequency/voltage with the neighboring generator it has communication with, (b) restoring its frequency/voltage to the nominal value, and (3) alter load sharing based on droop characteristics. If we denote $\epsilon_{v_k}$ and $\epsilon_{\omega_k}$ as voltage and frequency tracking errors for the $k^{\text{th}}$ generator, basing on (22) and (24) the following equations can be derived,

$$\dot{\epsilon}_{v_k} = \sum_{j=1}^{m} A_{\text{adj}(j,k)}(V_j - V_k) + \Psi(k,k)C_{gv(k,1)}(V_{ref} - V_k),$$
$$\dot{\epsilon}_{\omega_k} = \sum_{j=1}^{m} A_{\text{adj}(j,k)}(\omega_j - \omega_k) + \Psi(k,k)C_{g\omega(k,1)}(\omega_{ref} - \omega_k)$$
$$+ \Psi_{(k,k)}C_P \sum_{j=1}^{m} A_{\text{adj}(j,k)}\left(m_{p_k}P_{G_k} - m_{p_j}P_{G_j}\right), \quad (25)$$

We know that in complex network theory [21], the *Laplacian* matrix $\mathcal{L}$ is defined as

$$\mathcal{L} = \mathcal{D} - A_{\text{adj}}, \quad (26)$$

where $\mathcal{D}$ is the degree or valency matrix and $\mathcal{D}_{(i,i)} = \deg(v_i)$ and $\mathcal{D}_{(i,j)} = 0 (i \neq j)$ ($v_i \in \mathcal{N}$ is a node inside the node set of a graph $\mathcal{G}$), and the adjacency matrix is given as [21],

$$A_{\text{adj}(i,j)} = \begin{cases} 1 & \text{iff pair}\{v_i, v_j\} \in \mathcal{E}, \\ 0 & \text{otherwise} \end{cases} \quad (27)$$

where $\mathcal{E}$ denotes the edge set of a particular graph $\mathcal{G}$. Then the voltage and frequency tracking error in (25) for all DGs can be manipulated and expressed as,

$$\dot{\epsilon}_v = (\mathcal{L} + C_{gv}\Psi)\Xi_v,$$
$$\dot{\epsilon}_\omega = (\mathcal{L} + C_{g\omega}\Psi)\Xi_\omega, \quad (28)$$

where $\Xi_v$ and $\Xi_\omega$ are respectively the aggregation of voltage and frequency (including power sharing) control errors. Based on (28) and (25), we now have

$$\dot{\epsilon}_v = -C_v(\mathcal{L} + C_{gv}\Psi)\epsilon_v,$$
$$\dot{\epsilon}_\omega = -C_\omega(\mathcal{L} + C_{g\omega}\Psi)\epsilon_\omega, \quad (29)$$

which can be generalized by

$$\dot{\epsilon} = -G_c(\mathcal{L} + \mathcal{C}\Psi)\epsilon, \quad (30)$$

with $G_c$ representing the control gain, $\mathcal{C}$ the pinning gain and $\epsilon$ the control error.

*Lyapunov Stability:* A direct Lyapunov candidate $L$ can be assigned as,

$$L = \frac{1}{2}\epsilon^T\epsilon, \quad (31)$$

and apparently $L > 0$. Then we have

$$\dot{L} = \frac{1}{2}\left(\epsilon\dot{\epsilon}^T + \epsilon^T\dot{\epsilon}\right)$$
$$= \epsilon^T\dot{\epsilon} = \epsilon^T(-G_c(\mathcal{L} + \mathcal{C}\Psi)\epsilon)$$
$$\leq -G_c(\mathcal{L} + \mathcal{C}\Psi)||\epsilon||^2$$
$$< 0 \quad (32)$$

with a positive $G_c$ and appropriately chosen pinning gain $\mathcal{C}$ in correspondence to the *Laplacian* matrix and pinning matrix, hence Lyapunov stability proved. Despite the fact that Lyapunov stability is only a mild requirement for stability around the equilibrium point, it is assumed adequate for this study as the fluctuations of frequencies and voltages will be a small value under power system loading disturbances.

We are now in the position to look at the objective function that needs to be solved for the pinning-based PDMA problem. From (30) we know that the robustness and performance of the graph system are determined by the eigenvalues of $\mathcal{L} + \mathcal{C}\Psi$, such as the convergence rates of voltage and frequency, which is of paramount interest in this study, and can be quantitatively represented as $G_c\lambda_{\min}\{\mathcal{L} + \mathcal{C}\Psi\}$.

In this study, the main purpose of the pinning-based PDMA is to select a set of minimum number of generators and identify their identities so that the desired network convergence rate $\varrho^*$ can be achieved. To simplify this study, we assume the pinning gains are identical for frequency and voltage control, i.e., $C_{gv} = C_{g\omega}$ and $C_v = C_\omega = C_P$. Then our problem can be formulated as follows,

$$\begin{aligned}\text{minimize } &\mathcal{R}\{\Psi\},\\ \text{subject to } &G_c\lambda_{\min}\{\mathcal{L} + \mathcal{C}\Psi\} \geq \varrho^*,\end{aligned} \quad (33)$$

where $\mathcal{R}$ is the rank operator, and then with the minimum rank $\Upsilon_{\Psi\min}$ acquired from solving (33), we need to find which DGs to be pinned to realize the desired control purpose, i.e.,

$$\begin{aligned}\text{maximize } &\lambda_{\min}\{\mathcal{L} + \mathcal{C}\Psi\},\\ \text{subject to } &\mathcal{R}\{\Psi\} = \Upsilon_{\Psi\min}.\end{aligned} \quad (34)$$

*C. Deep Learning-based Pinning Decision-Making Method*

As mentioned in the previous sub-section, the pinning decision in the proposed PDMA control method is made by solving problems (33) and (34). However, solving such problem is a time consuming task, jeopardizing the applicability of the PDMA method to microgrids with time-vary communication networks which require real-time pinning set update. To overcome this limitation, we proposed a deep learning-based pinning decision-making method to obtain the optimal pinning set for real-time operations. First, a Monte Carlo simulation is conducted to generate a training set that covers a wide range of communication network configurations with different connection topologies. The details of the training data generation process is illustrated in the flowchart shown in Fig. 4, where GA is employed to obtain the optimal pinning sets. After the training set is generated, a deep neural network (DNN) is trained to capture the relationship between each topology and the corresponding optimal pinning set obtained by GA. The vectorized *Laplacian* matrix $\mathcal{L}_V = \text{vec}(\mathcal{L})$ and the optimal pinning set $\Psi_{\min}$ are employed as the input and the output of the DNN respectively. If undirected graphs are considered, the dimension of $\mathcal{L}_V$ can be reduced from $N^2$-by-1 to $((N+1)N/2)$-by-1 by removing repeated off-diagonal elements in the symmetrical *Laplacian* matrix.

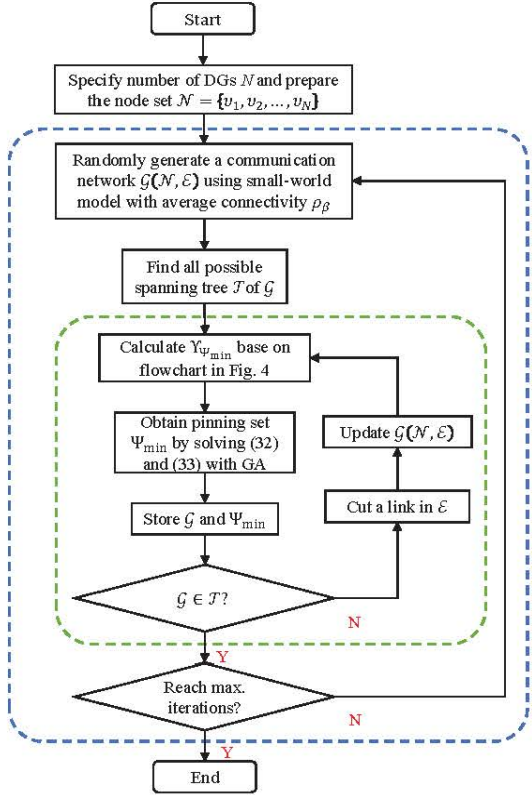

Fig. 4: Flowchart for training set generation process

## V. SIMULATION AND NUMERICAL RESULTS

In this study, a 38-node microgrid system is employed, which is modified from [22]. Fig. 5 demonstrates the system topology, where nodes 1 to 28 are the load nodes and whereas nodes 29 to 38 are the DG nodes. For simplicity but without loss of generality, all loads are modeled as $RL$ loads and the system operate in the islanded mode throughout the study. The load and generator settings are depicted in Fig. I. The line parameters and the inverter and LCL parameters are adopted and modified from [16] for the proposed new microgrid formation. The analysis and simulation are performed in MATLAB™ 2018b coding environment, and the time-domain simulation is performed by using the MATLAB™ built-in *ode15s* solver.

TABLE I: Load and Generator Settings

| Node | $m_p$ (p.u.) | $n_q$ (p.u.) | Node | $m_p$ | $n_q$ |
|---|---|---|---|---|---|
| 29 | $5.102 \times 10^{-3}$ | 0.02 | 34 | $5.102 \times 10^{-3}$ | 0.02 |
| 30 | $1.502 \times 10^{-3}$ | 1/30 | 35 | $1.502 \times 10^{-3}$ | 1/30 |
| 31 | $4.506 \times 10^{-3}$ | 0.02 | 36 | $4.506 \times 10^{-3}$ | 0.02 |
| 32 | $2.253 \times 10^{-3}$ | 0.05 | 37 | $2.253 \times 10^{-3}$ | 0.05 |
| 33 | $2.253 \times 10^{-3}$ | 0.05 | 38 | $2.253 \times 10^{-3}$ | 0.05 |

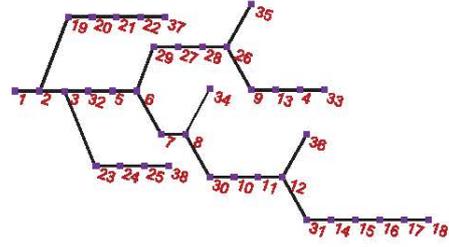

Fig. 5: Modified 38-node test system

### A. Power flow analysis

Power flow analysis is the first step to understand and observe a power system, which is also an essential step for the modal analysis and time-domain simulations in calculating the steady-state operating points and initial conditions. In this work, a novel microgrid power flow approach, as stated in Section III-A, is employed, and the power flow results are shown in TABLE II. In the microgrid of interest, node 29 is selected as the reference node, and its voltage angle is always assumed to be $0^o$. The secondary voltage and frequency control mechanism is implemented on all DG inverters. The nominal frequency and voltage of this microgrid are both 1 p.u. for the purpose of this study. In real-world situations the nominal values are generated by the tertiary control level, which is not incorporated in this paper. The secondary control is implemented to maintain this nominal voltage frequency by varying the no-load voltage and frequency of the DG droop characteristics, i.e., $V^{nl}$ and $\omega^{nl}$, as illustratively shown in Fig. 2(a). As shown in TABLE II, the no-load frequency of all DGs in the power flow problem converge to a single value, $\omega^{nl} = 1.0008$ p.u., whereas different no-load voltages are obtained for the DGs, ranging from 0.985 p.u. to 1.0387 p.u. This is because constraint (17) is incorporated to maintain the power sharing within the microgrid with respect to the capacity of each DG. Reactive power sharing is, however, not in the consideration in this study.

### B. Monte Carlo Simulation and DNN Model Training

In this subsection, a Monte Carlo simulation is conducted to generate the training data as mentioned in Section IV-C. The mircogrid system in this study consists of ten DGs, the communication network among each DG are modeled as a small-world network, with an average connectivity of 4. We assume all communication links are bidirectional, i.e., the measurement information is mutually exchanged between two DGs connected with a link connected with a link. Base on this assumption, formulas for undirected graph are used for the calculation of the degree matrix $\mathcal{D}$ and the *Laplacian* matrix $\mathcal{L}$ for each sample in this simulation. The *Laplacian* matrix of each sample is vectorized in order to generate the input for the DNN to be trained. In this case, the vectorized *Laplacian* matrix $\mathcal{L}_V$ is a 55-by-1 vector after removing the repeated off-diagonal entries of $\mathcal{L}$. By following the steps shown in Fig. 4, the training process is conducted. In this study, the DNN is trained with TensorFlow™ and Keras™ libraries using the Google™ Colab™ environment. The trained model is applied in the time-domain simulation for real-time optimal pinning set

TABLE II: Load flow results

| Node | Voltage $V$(p.u.) | $\theta(°)$ | Generation(-)/Load(+) $P$(kW) | $Q$(kVAr) | NL voltage $V^{nl}$(p.u.) |
|---|---|---|---|---|---|
| 1 | 0.9975 | -0.57 | -- | -- | -- |
| 2 | 0.9975 | -0.57 | 0.0989 | 0.0571 | -- |
| 3 | 0.9982 | -0.59 | 0.0898 | 0.0373 | -- |
| 4 | 0.9993 | 1.22 | 0.1181 | 0.0772 | -- |
| 5 | 1.0002 | -0.40 | 0.0595 | 0.0288 | -- |
| 6 | 0.9995 | -0.05 | 0.0599 | 0.0188 | -- |
| 7 | 0.9981 | -0.01 | 0.1965 | 0.0961 | -- |
| 8 | 0.9988 | 0.07 | 0.1967 | 0.0963 | -- |
| 9 | 0.9987 | 0.98 | 0.0599 | 0.0187 | -- |
| 10 | 0.9995 | 0.06 | 0.0591 | 0.0193 | -- |
| 11 | 0.9993 | 0.04 | 0.0443 | 0.0289 | -- |
| 12 | 0.9993 | 0.01 | 0.0594 | 0.0335 | -- |
| 13 | 0.9991 | 1.19 | 0.0590 | 0.0337 | -- |
| 14 | 0.9980 | -0.36 | 0.1187 | 0.0762 | -- |
| 15 | 0.9967 | -0.39 | 0.0588 | 0.0096 | -- |
| 16 | 0.9955 | -0.42 | 0.0598 | 0.0183 | -- |
| 17 | 0.9936 | -0.48 | 0.0585 | 0.0189 | -- |
| 18 | 0.9931 | -0.49 | 0.0897 | 0.0361 | -- |
| 19 | 0.9974 | -0.57 | 0.0890 | 0.0380 | -- |
| 20 | 0.9974 | -0.50 | 0.0883 | 0.0383 | -- |
| 21 | 0.9977 | -0.46 | 0.0898 | 0.0372 | -- |
| 22 | 0.9988 | -0.38 | 0.0891 | 0.0382 | -- |
| 23 | 0.9970 | -0.69 | 0.0883 | 0.0478 | -- |
| 24 | 0.9954 | -0.90 | 0.4109 | 0.1903 | -- |
| 25 | 0.9969 | -1.07 | 0.4118 | 0.1913 | -- |
| 26 | 0.9994 | 0.81 | 0.0590 | 0.0241 | -- |
| 27 | 1.0001 | 0.09 | 0.0599 | 0.0236 | -- |
| 28 | 0.9994 | 0.49 | 0.0591 | 0.0193 | -- |
| 29 | 1.0000 | 0.00 | -0.1494 | -0.3170 | 1.0063 |
| 30 | 1.0000 | 0.20 | -0.5076 | 0.3236 | 0.9892 |
| 31 | 1.0000 | -0.29 | -0.1692 | -0.5505 | 1.0110 |
| 32 | 1.0000 | -0.54 | -0.3384 | -0.7747 | 1.0387 |
| 33 | 1.0000 | 1.29 | -0.3384 | 0.0157 | 0.9992 |
| 34 | 1.0000 | 0.22 | -0.1494 | 0.0497 | 0.9990 |
| 35 | 1.0000 | 1.49 | -0.5076 | 0.4505 | 0.9850 |
| 36 | 1.0000 | 0.20 | -0.1692 | 0.1080 | 0.9978 |
| 37 | 1.0000 | -0.33 | -0.3384 | -0.0416 | 1.0021 |
| 38 | 1.0000 | -1.12 | -0.3384 | -0.6482 | 1.0324 |
| System frequency $\omega$ : 1 p.u. | | | No load frequency $\omega^{nl}$ : 1.0008 p.u. | | |

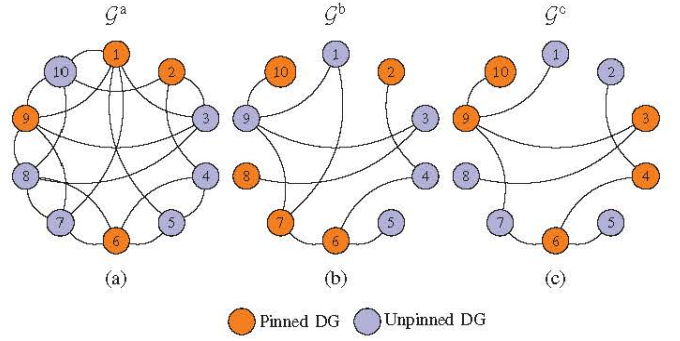

Fig. 6: Communication network under stochastic disruptions

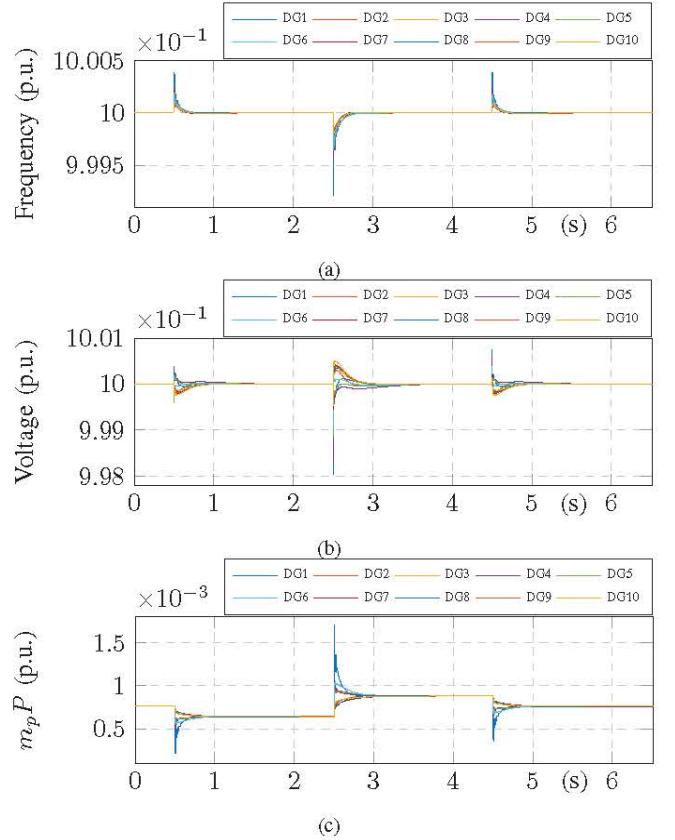

Fig. 7: DGs' (a) frequencies, (b) voltages and (c) real power sharing

generation, which will be discussed in detail in the following subsection.

### C. Time-Domain Simulation

In this subsection, time-domain simulations are carried out on the modified 38-node microgrid under the islanded mode. The DNN model trained in the previous subsection is imported to MATLAB™ for generation of pinning sets used in the proposed PDMA control method. The microgrid system operating in steady-state for the first 0.5s, and disturbances are then introduced to the microgrid in order to verify the proposed PDMA method. The time-domain simulation results will further illustrate the control performance in the case study.

*1) Case 1: PDMA with fixed communication network:* In this case, the communication network is represented by $\mathcal{G}^a$ as shown in Fig. 6(a), which used to verify the functionality of the proposed PDMA control method. Given the gain of the frequency, voltage and real power sharing controller both equal to $c_\omega = c_V = c_p = 30$ and the desired convergence rate $\varrho^* = 10$, the desired algebraic connectivity to the reference is found to be $\mu^*_{\omega,V} \geq \varrho^*/C_{\omega,V} = 1/3$, and the resultant pinning set is $\Psi^a = \{DG1, DG2, DG6, DG9\}$ as highlighted in orange in Fig. 6(a), with the calculated minimum network convergence rate $G_c\lambda_{\min}(\Psi^a) = 10.8318 > \varrho^*$. The time-domain simulation runs for 6.5 seconds with the microgrid operates in the steady-state for the first 0.5 seconds, an abrupt load decrease is then introduced at $t = 0.5s$, with a 50% reduction on the load demand at nodes 2 to 10. After that, the load demand of these nodes is increased by 100%, and the loading condition is returned to the original at $t = 4.5$. Note that the cyber network remains unchanged throughout this case study. Time-domain simulation results in Fig. 7 further confirm that the frequencies and voltages are restored to the nominal values, and real power sharing among the DGs reaches consensus after the load variation disturbances. The results also demonstrated that the proposed method is capable to maintain autonomous operations of the microgrid through droop-based secondary control strategy.

*2) Case 2: PDMA with communication network under stochastic disruptions:* In this case, stochastic communication network disruptions are simulated to study the transient behaviors of the DGs in the microgrid as well as the effectiveness of the proposed pinning decision making method. As mentioned in Section IV-C, the pinning sets used in the PDMA control are updated continuously in order to keep the network convergence rate below the desired value $\varrho^* = 10$ using the results obtained based on the pinning decision making method. In this particular study, GA is used to generate the pinning sets due to the fact that the size of the microgrid is small. Note that the GA-based decision making method is not applicable when the size of the microgrid is grown, and the DL-based decision-making method will be used as discussed in Section IV-C. To simulate the stochastic disruptions, communication links in the network are taken out randomly as shown in Fig. 6, where the original cyber network is illustrated in Fig. 6(a). First, edges $\{v_1, v_3\}$, $\{v_1, v_5\}$, $\{v_1, v_{10}\}$, $\{v_2, v_3\}$, $\{v_2, v_{10}\}$, $\{v_4, v_5\}$, $\{v_6, v_8\}$ are removed from Fig. 6(a), and the resulting network connection $\mathcal{G}^b$ is shown in Fig. 6(b). After the disruption, the network convergence rate with the original pinning set has dropped from 10.8318 to 8.4479, which violates the design specification. Therefore, to fulfill the network convergence rate requirement, a new pinning set is obtained with GA using the proposed method, with $\Psi^b = \{DG2, DG6, DG7, DG8, DG10\}$ and $\varrho^b = 11.7871$. Note that the number of pinned DGs has increased to $|\Psi^b| = 5$ after the disruption, i.e., pinning four DGs can no longer produce a result that satisfies the design specification. Similarly, an additional link is disconnected from DG1 and DG7 in order to further examine the proposed method, and the pinning set is updated again using the same method. As shown in Fig. 6(c), the new pinning set for $\mathcal{G}^b$ has the same cardinality with the previous pinning set, i.e., $|\Psi^b| = |\Psi^c| = 5$, but DG3, DG4, DG6, DG9 and DG10 are pinned instead, and the network convergence rate is calculated to be $\varrho^c = 11.8210$.

## VI. Conclusion

In this study, we have proposed a pinning strategy by viewing the power system as a cyber-physical system, for PDMA-AGC in microgrids. The proposed pinning method aims to make adaptive pinning decisions with the minimum number pinned generators against communication disruptions between control agents, to achieve frequency consensus amongst all generators and restore the power system frequency to the nominal value. A DNN-based learning technique is employed to make online pinning decisions upon cyber-system topology changes. Time-domain simulations on the modified IEEE 10-generator, 38-bus microgrid test system have verified the proposed pinning decision-making approach. Future work may include testing and verification on large-scale microgrid systems and further investigations on the correlation between the physical and cyber components in complex power systems.